\documentclass[a4paper,12pt]{article}
\usepackage[pctex32]{graphics}
\usepackage{amssymb,amsmath}
\usepackage{amsmath,amssymb}
\usepackage{latexsym}
\usepackage{epsfig}
\usepackage[english]{babel}

\begin{document}

\begin{titlepage}

\vspace{5mm}
\begin{center}
{\Large \bf  Superradiant instability of charged massive scalar field
 in Kerr-Newman-anti-de Sitter black hole}

\vskip 1.6cm

\centerline{\large Ran Li$^*$}

\vskip .6cm

{Department of Physics, Henan Normal University, Xinxiang 453007, China \\}

\vskip 1.6cm

\end{center}

\begin{center}

\underline{\large Abstract}
\end{center}
 We study the superradiance instability of charged massive scalar field in
 the background of Kerr-Newman-anti-de Sitter black hole. By employing
 the asymptotic matching technique to solve Klein-Gordon equation analytically,
 the complex parts of quasinormal frequencies are shown to be positive
 in the regime of superradiance. Because the calculation is performed
 in the approximation of small black hole, the result indicates that
 small Kerr-Newman-anti-de Sitter black hole is unstable against the
 massive scalar field perturbation with small charge.

\vskip 1.6cm

\noindent Keywords: Kerr-Newman-anti-de Sitter black hole, superradiant instability
\vskip 0.8cm

\vspace{15pt} \baselineskip=18pt

\vskip 5.6cm

\noindent $^*$Electronic address: liran.gm.1983@gmail.com

\thispagestyle{empty}
\end{titlepage}

\newpage
\section{Introduction}

 In an rotating charged black hole with ergosphere outside the event horizon,
 the impinging bosonic fields with frequency $\omega$, angular
 momentum $m$, and electric charge $e$ inside a specific band
 \begin{eqnarray}
 0<\omega<m\Omega_H+e\Phi_H\;,
 \end{eqnarray}
 where $\Omega_H$ is angular velocity of event horizon, and $\Phi_H$
 is electrostatic potential at horizon,
 can be scattered off with a higher amplitude. This phenomenon,
 which is called superradiance \cite{zeldovich,bardeen,misner,starobinsky},
 allows extracting rotational energy and electrostatic energy efficiently
 from black hole.

 It was suggested by Press and Teukolsky \cite{press} to use superradiance phenomenon
 to built a black-hole bomb. If one add a reflecting mirror outside the
 rotating black hole, superradiant modes will bounce back and forth between
 black hole and mirror and the energy extracted from black hole
 will grow exponentially. This mechanism has been recently restudied by many authors
 \cite{cardoso2004bomb,Hod,Rosa,Lee}.

 However, the reflecting mirror is not necessarily artificial. In fact,
 nature can provide its own mirror sometimes. In these cases, black holes
 are unstable due to the superradiance, which is called superradiant instability.
 For example, the asymptotically flat Kerr and Kerr-Newman black holes
 have this kind of instability against massive scalar field perturbations,
 where the effective reflective mirror is sourced from mass terms \cite{kerrunstable,detweiler,dolan,hodPLB2012}.
 Another example investigated by Konoplya \cite{konoplyaPLB} is that
 the Kerr-Newman black hole immersed in a asymptotically uniform magnetic field also
 shows superradiant instability for massive scalar perturbation. The boosted
 Kerr black string considered by Dias \cite{DiasPRD2006} is also unstable due to superradiance
 of massless scalar field where Kaluza-Klein momentum along
 the string direction works as effective reflective mirror.

 For the rotating black hole in anti-de Sitter (AdS) space, the boundary of AdS spacetime
 behaves as reflecting mirror and superradiant modes can be bounded
 in the effective potential barrier, which leads to instability of black hole.
 It is found by Cardoso and Dias \cite{cardoso2004ads} that the small Kerr-AdS black hole is unstable
 against massless scalar field perturbation. Latter, superradiant instability
 of small Kerr-AdS black hole against gravitational perturbation is also established \cite{cardoso2006prd}.
 For the five-dimensional rotating charged AdS black hole \cite{aliev}, only the superradiant modes
 with even orbital quantum number can trigger the instability, which is contrary
 to the case of four-dimensional Kerr-AdS black hole. More recently, superradiant
 instability of small Reissner-Nordstr\"{o}m-anti-de Sitter black hole
 is investigated analytically and numerically \cite{uchikata}. In fact, there are also other cases
 that the boundary of spacetime can provide reflecting mirror. For example,
 the rotating linear dilaton black hole \cite{clement} and the Myers-Perry black hole in G\"{o}del
 universe \cite{knopolya} are unstable due to the Dirichlet boundary condition at asymptotic infinity.

 As we have discussed in the above, superradiant instability naturally happens
 when two conditions are satisfied (one can refer to \cite{konoplya2011rmp}
 for a recent review on this aspect): (1) Black hole has rotation or charge;
 (2) There is a natural reflecting mirror outside the black hole.
 So it is natural to say
 that Kerr-Newman-anti-de Sitter (KN-AdS) black hole also have 
 superradiant instability. Although, as far as we know,
 an explicit computation of superradiant instability of KN-AdS black hole is still absent
 in literatures. In this paper, we will perform the calculation of superradiant
 instability of charged massive scalar field for KN-AdS black hole.
 The method we have used is just the so-called matched asymptotic expansion method.
 We divide the space outside the event horizon into near-region and far-region.
 The solution can be approximated by matching the near-region solution and the far-region solution
 in an overlapping region, which allows us to analysis the complex quasinormal modes by imposing the boundary conditions. The result shows that the complex parts of quasinormal frequencies are positive
 in the regime of superradiance. In order to apply the asymptotic matching method, 
 we have assumed that the energy of scalar field is 
 low and the charge of scalar field is small 
 (one can refer to Eq.(\ref{assumption}) for the exact formulation). 
 In addition, the calculation is performed
 for the small black hole case, where the size of 
 black hole event horizon is much smaller 
 than the AdS radius. The result indicates that
 the small KN-AdS black hole is unstable against the massive
 scalar perturbation with small charge.

 This paper is arranged as follows. In Section 2,
 we give a brief review of KN-AdS
 black hole. In Section 3, we investigate the superradiance
 and the boundary condition of charged massive scalar field.
 In Section 4, solution of Klein-Gordon equation is solved
 by using the matching technique and superradiant instability
 is explicitly shown. The last section is devoted to conclusion and discussion.

 \section{Kerr-Newman-anti-de Sitter black hole}

 The metric of KN-AdS black hole \cite{carter} in the Boyer-Lindquist-type coordinates
 takes the form
 \begin{eqnarray}\label{metric}
  ds^2&=&-\frac{\Delta_r}{\rho^2}\left[dt-\frac{a\sin^2\theta}{\Sigma}d \phi\right]^2
  +\frac{\rho^2}{\Delta_r}dr^2+\frac{\rho^2}{\Delta_\theta}d\theta^2
  \nonumber\\&&
  +\frac{\Delta_\theta\sin^2\theta}{\rho^2}\left[a dt-\frac{r^2+a^2}{\Sigma}d\phi\right]^2\;,
 \end{eqnarray}
 where
 \begin{eqnarray}
  \rho^2&=&r^2+a^2\sin^2\theta\;,\nonumber\\
  \Sigma&=&1-\frac{a^2}{L^2}\;,\nonumber\\
  \Delta_r&=&(r^2+a^2)\left(1+\frac{r^2}{L^2}\right)-2Mr+Q^2\;,\nonumber\\
  \Delta_\theta&=&1-\frac{a^2}{L^2}\cos^2\theta\;.
 \end{eqnarray}

  The three parameters $M$, $a$, and $Q$ are related to mass,
  angular momentum, and electric charge of black hole. This metric is a
  solution to Einstein-Maxwell equations with a negative cosmological
  constant $\Lambda=-3/L^2$. The electro-magnetic vector potential is given by
  \begin{eqnarray}
   A_\mu dx^\mu=-\frac{Qr}{\rho^2}\left(dt-\frac{a\sin^2\theta}{\Sigma}d\phi\right)\;.
  \end{eqnarray}

  In this paper, we will consider the non-extremal black hole case.
  The metric function $\Delta_r$ has two real roots $r_\pm$. The event
  horizon locates at the largest root $r_+$ of $\Delta_r$.
  For latter convenience, we give the expressions of angular
  velocity $\Omega_H$ and electric potential $\Phi_H$ at event horizon
  \begin{eqnarray}
   \Omega_H=\frac{a\Sigma}{r_+^2+a^2}\;,\;\;\;\Phi_H=\frac{Qr_+}{r_+^2+a^2}\;.
  \end{eqnarray}

  \section{Superradiance of the charged massive scalar field}

 Superradiance phenomenon of massive scalar field in KN-AdS black hole
 has been studied in \cite{win}. In this section, we will reconsider the classical
 superradiance of charged massive scalar field. The boundary conditions
 for solving the wave equation are also presented in this section.
 We start with the field equation for charged massive scalar field perturbation
 which is given by Klein-Gordon equation
 \begin{eqnarray}
  \left(\nabla_\mu-ieA_\mu\right)\left(\nabla^\mu-ieA^\mu\right)\Phi=\mu^2\Phi\;,
 \end{eqnarray}
 where $e$ and $\mu$ are charge and mass of scalar field.
 Taking the ansatz
 \begin{eqnarray}
  \Phi=e^{-i\omega t+im\phi}S(\theta)R(r)\;,
 \end{eqnarray}
 the wave equation can be separated into
 \begin{eqnarray}
 \Delta_r\partial_r(\Delta_r\partial_r R(r))&+
 &\left[(\omega(r^2+a^2)-ma\Sigma-eQr)^2\right.\nonumber\\
 &&\left.-\Delta_r(K+\mu^2 r^2)\right]R(r)=0\;,
 \end{eqnarray}
 \begin{eqnarray}
 \Delta_\theta\sin\theta\partial_\theta(\Delta_\theta\sin\theta\partial_\theta S(\theta))
 &+&\left[K\Delta_\theta\sin^2\theta-(\omega a\sin^2\theta-m\Sigma)^2
 \right.\nonumber\\&&\left.
 -\mu^2a^2\Delta_\theta\sin^2\theta\cos^2\theta\right]S(\theta)=0\;,
 \end{eqnarray}
 where $K$ is the constant of separation. In the next section, we will
 interest in the regime of $\omega a\ll 1$ and $a/L\ll 1$. In this paper,
 we also consider the case that the frequency of scalar field modes is nearly equal to
 the mass of scalar field, i.e. $\omega\sim\mu$. This implies that
 $\mu a\ll 1$. Using these assumptions, the angular equation can be reduced to
 the form of spherical harmonics equation
 \begin{eqnarray}
  \frac{1}{\sin\theta}\partial_\theta(\sin\theta S(\theta))
  +\left(K-\frac{m^2}{\sin^2\theta}\right)S(\theta)=0\;.
 \end{eqnarray}
 It is obvious that the separation constant $K$ is just $l(l+1)$.

 Defining the tortoise coordinate $r^*$ and a new radial function
 $\tilde{R}$ as
 \begin{eqnarray}
  \frac{dr^*}{dr}=\frac{r^2+a^2}{\Delta_r}\;,\;\;\;
  \tilde{R}(r^*)=R(r)(r^2+a^2)^{1/2}\;,
 \end{eqnarray}
 the radial function can be rewritten in the form of Schr\"{o}dinger equation
 \begin{eqnarray}
 \frac{d^2\tilde{R}(r^*)}{d r^{*2}}+V(r^*)\tilde{R}(r^*)=0\;,
 \end{eqnarray}
 where the effective potential $V(r^*)$ is explicitly given by
  \begin{eqnarray}
 V(r^*)&=&\left(\omega-\frac{ma\Sigma}{r^2+a^2}-\frac{eQr}{r^2+a^2}\right)^2
 -\frac{\Delta_r(K+\mu^2 r^2)}{(r^2+a^2)^2}+\frac{3r^2\Delta_r^2}{(r^2+a^2)^4}
 \nonumber\\
 &&-\frac{\Delta_r}{(r^2+a^2)^3}\left(3r^2+a^2+\frac{5r^4}{L^2}-\frac{3a^2r^2}{L^2}
 +4Mr-Q^2\right)\;.
 \end{eqnarray}

 Near the horizon $r_+$, where $\Delta_r\rightarrow 0$, the effective potential
 can be approximated as
 \begin{eqnarray}
 V(r^*)\rightarrow (\omega-m\Omega_H-e\Phi_H)^2\;.
 \end{eqnarray}
 Because the classical wave is considered, the ingoing boundary
 condition at the event horizon should be employed. The ingoing field modes are given by
 \begin{eqnarray}
 \Phi\sim e^{-i\omega t-i(\omega-m\Omega_H-e\Phi_H) r^*}\;.
 \end{eqnarray}
 This is the boundary condition of radial wave equation at the event horizon.

 From Eq.(15), one can also obtain the superradiance condition which is given by
 \begin{eqnarray}
 0<\omega<m\Omega_H+e\Phi_H\;.
 \end{eqnarray}
 It can be easily shown that the amplitude of a wave will be amplified after
 scattering by the event horizon when the frequency is in the superradiant regime.

 Near the infinity, $r\rightarrow \infty$, the effective potential can be
 approximated as
 \begin{eqnarray}
 V(r^*)\rightarrow -\left(\frac{\mu^2}{L^2}+\frac{2}{L^4}\right)r^2
 \end{eqnarray}
 One can see that the mass of scalar field as well as the cosmological constant term
 provide the natural infinity high potential that can leads to superradiant instability.
 We require the boundary condition near the asymptotic infinity is just the 
 Dirichlet boundary condition
 \begin{eqnarray}
 \Phi\rightarrow 0\;,\;\;\; \textrm{as}\;\;\; r\rightarrow \infty\;.
 \end{eqnarray}

 With the boundary conditions that the ingoing wave at the horizon and the Dirichlet boundary condition at the asymptotic infinity, one can solve the complex quasinormal modes of charged massive scalar field
 in KN-AdS background. If the imaginary part of quasinormal mode is negative, it is known that
 the system is stable against this kind of perturbation. The instability means that the imaginary
 part is positive. In the next section, we will calculate the quasinormal modes by using the asymptotic
 matching technique. It is shown that, in the regime of superradiance, the imaginary part
 of quasinormal mode is positive. In other words, superradiant instability of KN-AdS black hole is found analytically.

 \section{Analytical calculation of superradiant instability}

 In this section, we will present an analytical calculation of
 superradiant instability for charged massive scalar perturbation.
 In order to employ the matched asymptotic expansion method, we need to
 take the assumption that the parameters satisfy the conditions 
 \begin{eqnarray}\label{assumption}
 O(\omega M)=O(qQ)\equiv O(\epsilon), \;\;\; \epsilon\ll 1.  
 \end{eqnarray}
 The angular momentum of black hole is assumed to satisfy $O(a/M)=1$.
 Then we can divide
 the space outside the event horizon into two regions, namely, a near-region,
 $r-r_+\ll 1/\omega$, and a far-region, $r-r_+\gg M$. The approximated solution
 can be obtained by matching the near-region solution and the far-region solution
 in the overlapping region $M\ll r-r_+\ll1/\omega$. By imposing the boundary conditions,
 we can analysis the properties of the solution and study the stability of black hole
 against the perturbation.

 \subsection{Near-region solution}

 Let us firstly focus on the near-region in the vicinity of the horizon,
 $\omega(r-r_+)\ll 1$, and the regime of low frequency perturbation, $\omega r_+\ll 1$.
 The low frequency perturbation also implies that $\mu^2 r^2\ll 1$ in the near-region.
 For the small AdS black holes, $r_+/L\ll 1$,
 the metric function can be approximated as
 \begin{eqnarray}
 \Delta_r\cong\Delta=r^2-2Mr+a^2+Q^2=(r-r_+)(r-r_-)\;.
 \end{eqnarray}
 Then, the radial wave function in the near-region can be reduced to the form
 \begin{eqnarray}
 \Delta\partial_r(\Delta\partial_rR(r))
 +\left[(\omega(r_+^2+a^2)-ma\Sigma-eQr_+)^2
 \right.\nonumber\\\left.
 -l(l+1)\Delta\right]R(r)=0\;.
 \end{eqnarray}
 Introducing the coordinate
 \begin{eqnarray}
 z=\frac{r-r_+}{r-r_-}\;,
 \end{eqnarray}
 the near-region radial equation can be written in the form of
 \begin{eqnarray}
 z\frac{d}{dz}\left(z\frac{d}{dz}R(z)\right)
 +\left[\varpi^2-l(l+1)\frac{z}{(1-z)^2}\right]R(z)=0\;,
 \end{eqnarray}
 with
  \begin{eqnarray}
 \varpi=\frac{r_+^2+a^2}{r_+-r_-}\left(\omega
 -\frac{ma\Sigma}{r_+^2+a^2}-\frac{eQr_+}{r_+^2+a^2}\right)\;.
 \end{eqnarray}
 The near-region solution with the ingoing boundary condition is given by
 \begin{eqnarray}
 R(z)=Az^{-i\varpi}(1-z)^{l+1}F(l+1,l+1-2i\varpi,1-2i\varpi,z)\;.
 \end{eqnarray}
 Using the $z\rightarrow 1-z$ transformation law for the hypergeometric function,
 one can get the large $r$ behavior of the near-region solution as
  \begin{eqnarray}
 R&\sim& A\Gamma(1-2i\varpi)\left[\frac{(r_+-r_-)^{-l}\Gamma(2l+1)}
 {\Gamma(l+1)\Gamma(l+1-2i\varpi)}r^l\right.
 \nonumber\\&&\left.
 +\frac{(r_+-r_-)^{l+1}\Gamma(-2l-1)}
 {\Gamma(-l)\Gamma(-l-2i\varpi)}r^{-l-1}\right]\;.
 \end{eqnarray}
 This solution should be matched with the small $r$ behavior of the far-region solution
 obtained in the next subsection.

 \subsection{Far-region solution}

 In the Far-region, $r-r_+\gg M$, we can neglect the effects induced by the black hole,
 i.e. we have $a\sim 0$, $M\sim 0$, $Q\sim 0$, and $\Delta_r\cong r^2(1+r^2/L^2)$.
 One can deduce the far-region wave equation as
 \begin{eqnarray}
 \left(1+\frac{r^2}{L^2}\right)\partial_r^2R(r)&+&
 2r\left(\frac{2}{L^2}+\frac{1}{r^2}\right)\partial_r R(r)
 \nonumber\\&+&\left[\frac{\omega^2}{1+r^2/L^2}
  -\frac{l(l+1)}{r^2}
 -\mu^2\right]R(r)=0\;.
 \end{eqnarray}

 Introducing the new radial coordinate as
 \begin{eqnarray}
  x=1+\frac{r^2}{L^2}\;,
 \end{eqnarray}
 the far-region radial equation can be rewritten as
 \begin{eqnarray}
 x(1-x)\partial_x^2 R&+&\left(1-\frac{5}{2}x\right)\partial_x R
 \nonumber\\
 &-&
 \left[\frac{w^2L^2}{4x}+\frac{l(l+1)}{4(1-x)}-\frac{\mu^2L^2}{4}\right]R=0\;.
 \end{eqnarray}
 The solution is given by the hypergeometric function
 \begin{eqnarray}
 R(x)=\left(1/x-1\right)^{-\frac{l+1}{2}}\left[
 C_1x^{-\frac{3}{4}-\frac{1}{2}(\alpha-\beta)}F(\alpha,\alpha-\gamma+1,
 \alpha-\beta+1,1/x)\right.\nonumber\\+\left.
 C_2x^{-\frac{3}{4}+\frac{1}{2}(\alpha-\beta)}F(\beta,\beta-\gamma+1,
 \beta-\alpha+1,1/x)\right]\;,
 \end{eqnarray}
 with the parameters given by
 \begin{eqnarray}
 \alpha&=&-\frac{l}{2}+\frac{1}{4}+\frac{1}{4}\sqrt{9+4\mu^2L^2}+\frac{\omega L}{2}\;,
 \nonumber\\
 \beta&=&-\frac{l}{2}+\frac{1}{4}-\frac{1}{4}\sqrt{9+4\mu^2L^2}+\frac{\omega L}{2}\;,
 \nonumber\\
 \gamma&=&1+\omega L\;.
 \end{eqnarray}

 When $x\rightarrow \infty$, by using the property of hypergeometric
 function $F(a,b,c,0)=1$, one can arrive at the asymptotic behavior of
 the far-region solution
 \begin{eqnarray}
 R\rightarrow C_1x^{-\frac{3}{4}-\frac{1}{4}\sqrt{9+4\mu^2L^2}}
 +C_2x^{-\frac{3}{4}+\frac{1}{4}\sqrt{9+4\mu^2L^2}}\;.
 \end{eqnarray}
 According to the Dirichlet boundary condition at infinity, one must set $C_2=0$.
 Then the far-region solution is given by
 \begin{eqnarray}
 R(x)=C_1\left(1-x\right)^{-\frac{l+1}{2}}x^{\frac{l}{2}-\frac{1}{4}-\frac{1}{2}(\alpha-\beta)}
 F(\alpha,\alpha-\gamma+1,\alpha-\beta+1,1/x)
 \end{eqnarray}

 In order to match the near-region solution, we must study the small $r$ behavior
 of the far-region solution. By using the $\frac{1}{x}\rightarrow 1-x$ transformation property
 of hypergeometric function, we have the following formula when $x\rightarrow 1$
 \begin{eqnarray}
  F(\alpha,\alpha-\gamma+1,\alpha-\beta+1,1/x)&\cong&
  (x-1)^{\gamma-\alpha-\beta}\frac{\Gamma(\alpha-\beta+1)\Gamma(\alpha+\beta-\gamma)}
  {\Gamma(\alpha)\Gamma(\alpha-\gamma+1)}
  \nonumber\\
  &&+\frac{\Gamma(\alpha-\beta+1)\Gamma(\gamma-\alpha-\beta)}
  {\Gamma(1-\beta)\Gamma(\gamma-\beta)}\;.
 \end{eqnarray}
 Then, in the small $r$ limit, the far-region solution is given by
 \begin{eqnarray}
  R(r)\sim C_1(-1)^{(l+1)/2}\Gamma(\alpha-\beta+1)\left[\frac{\Gamma(\alpha+\beta-\gamma)}
  {\Gamma(\alpha)\Gamma(\alpha-\gamma+1)}\left(\frac{r}{L}\right)^l
  \right.\nonumber\\
  \left.+\frac{\Gamma(\gamma-\alpha-\beta)}
  {\Gamma(1-\beta)\Gamma(\gamma-\beta)}\left(\frac{r}{L}\right)^{-l-1}\right]\;.
 \end{eqnarray}

  The requirement of the regularity of the wave solution at the origin
  selects the frequencies of massive scalar field that might propagate in the
  AdS background. In order to have a regular solution at the origin we must
  demand that $\Gamma(1-\beta)=\infty$. So we have
  \begin{eqnarray}
   \omega_n=\frac{3+\sqrt{9+4\mu^2L^2}+2l+4n}{2L}\;.
  \end{eqnarray}

  We assume that the frequencies have a small imaginary part $\delta$
  which is induced by black hole event horizon
  \begin{eqnarray}
  \omega=\omega_n+i\delta\;.
  \end{eqnarray}
  Then, the far-region solution in the small $r$ region can be approximated
  as
  \begin{eqnarray}
   R(r)\sim C_1(-1)^{(l+1)/2}
   \left[\frac{\Gamma(1+\sqrt{9+4\mu^2L^2}/2)\Gamma(-l-1/2)}
  {\Gamma(n+1+\sqrt{9+4\mu^2L^2}/2)\Gamma(-l-1/2-n)}\left(\frac{r}{L}\right)^l
  \right.\nonumber\\
  \left.+(-1)^{n+1}n!\left(\frac{\delta L}{2}\right)\frac{\Gamma(1+\sqrt{9+4\mu^2L^2}/2)\Gamma(l+1/2)}
  {\Gamma(n+l+3/2+\sqrt{9+4\mu^2L^2}/2)}\left(\frac{r}{L}\right)^{-l-1}\right]\;,
  \end{eqnarray}
  where we have used the formula
  \begin{eqnarray}
  \frac{1}{\Gamma(-n-i\delta L/2)}=(-1)^{n+1}n!i\delta L/2\;.
  \end{eqnarray}

  \subsection{Matching condition: the unstable modes}

  The matching of the near-region solution in the large $r$ region and
  the far-region solution in the small $r$ region yields
  \begin{eqnarray}
  \delta\cong -\sigma(\omega_n-m\Omega_H-e\Phi_H)
  \frac{(r_+^2+Q^2)(r_+-r_-)^{2l}}{L^{2l+2}}\;,
  \end{eqnarray}
  where
  \begin{eqnarray}
  \sigma&=&\frac{2^{l-n+2}}{\sqrt{\pi}}
  \frac{(l!)^2(2l+1+n)!!}{n!(2l+1)!(2l)!(2l+1)!!(2l-1)!!}
  \nonumber\\&&\times\frac{\Gamma(n+l+3/2+\sqrt{9+4\mu^2L^2}/2)}
  {\Gamma(n+1+\sqrt{9+4\mu^2L^2}/2)}
  \left[\prod_{k=1}^{l}(k^2+4\varpi^2)\right]\;,
  \end{eqnarray}
  with $\varpi=(\omega_n-m\Omega_H-e\Phi_H)(r_+^2+Q^2)/(r_+-r_-)$.
  So, it is easy to see that, in the superradiance regime, $Re[\omega]-m\Omega_H-e\Phi_H<0$,
  the imaginary part of complex frequency
  $\delta>0$, which leads to the instability of these modes. In other words, we can conclude
  that the small KN-AdS black hole is unstable against the charged massive scalar field perturbation.
  This instability is caused by the superradiance of scalar field.

  \section{Conclusion and discussion}

 In summary, we have studied the instability of small KN-AdS black hole
 caused by the superradiance of scalar field. This kind of instability
 originates from the superradiance of boson field and the Dirichlet boundary
 condition. It is shown in Section 3, for the charged massive scalar field in KN-AdS
 black hole, the mass of scalar field as well as the cosmological constant term
 provide the natural infinity high potential at asymptotic infinity. This leads to the Dirichlet
 boundary condition at infinity. In other words, the mass term and the cosmological constant term
 play the roll of mirror that can reflect superradiant modes.

 We have analytically investigated this kind of instability by calculating
 the complex quasinormal modes for small KN-AdS black hole.
 It is explicitly shown that the imaginary part of complex quasinormal modes
 is positive in the regime of superradiance. The method we have employed 
 is the asymptotic expansion method. To apply this method, 
 we have made the assumption that the energy and the charge of scalar field 
 are small. Our result indicates that the small Kerr-Newman-anti-de Sitter black hole 
 is unstable against the massive scalar perturbation with small charge. 
 This conclusion is different from that of Reissner-Nordstr\"{o}m-AdS black 
 hole\footnote{The author thanks the anonymous referee for pointing this out.},
 where the instability is shown for the scalar field with the large charge \cite{maeda,abdalla}. 
 It is also shown that this instability results in the charge extraction from  
 Reissner-Nordstr\"{o}m-AdS black hole. So it will be interesting to study the 
 relation of instability of large charge scalar
 field in Reissner-Nordstr\"{o}m-AdS black hole and superradiance phenomenon.

 Recently, the superradiant instability is found for the charged Myers-Perry black hole
 in G\"{o}del universe by using numerical method \cite{knopolya}.
 It is shown that the spectrum of scalar field perturbation includes the
 superradiant unstable modes. It may be interesting to perform an analytical
 calculation to show the superradiant instability in this background
 explicitly by using the matching asymptotic expansion method.

 \section*{Acknowledgement}

 I would like to thank Pu-Jian Mao and Ming-Fan Li for reading the manuscript and useful
 comments. This work was supported by NSFC, China (Grant No. 11147145).

\end{document}